\documentclass[english,prd,twocolumn,superscriptaddress,nofootinbib,preprintnumbers]{revtex4}
\usepackage[latin1]{inputenc}
\usepackage{graphicx}
\usepackage{amssymb}

\makeatletter

\usepackage{amsfonts}
\usepackage{dcolumn}
\usepackage{hyperref}

\def\be{\begin{equation}}
\def\ee{\end{equation}}
\def\ba{\begin{eqnarray}}
\def\ea{\end{eqnarray}}
\def\bs{\begin{subequations}}
\def\es{\end{subequations}}

\newcommand{\rd}{{\rm d}}

\usepackage{color}

\makeatother

\usepackage{babel}
\makeatother
\begin{document}

\title{The effect of modified gravity on weak lensing}

\author{Shinji Tsujikawa}
\affiliation{Department of Physics, Faculty of Science, Tokyo University of Science, 
1-3, Kagurazaka, Shinjuku-ku, Tokyo 162-8601, Japan}
\email{shinji@rs.kagu.tus.ac.jp}

\author{Takayuki Tatekawa}
\affiliation{Department of Computer Science, Kogakuin University,
1-24-2 Nishi-shinjuku, Shinjuku, Tokyo, 163-8677 Japan}
\email{tatekawa@cpd.kogakuin.ac.jp}
\affiliation{Research Institute for Science and Engineering,
Waseda University,
3-4-1 Okubo, Shinjuku, Tokyo, 169-8555 Japan}

\begin{abstract}

We study the effect of modified gravity on weak lensing 
in a class of scalar-tensor theory that includes $f(R)$ gravity 
as a special case. 
These models are designed to 
satisfy local gravity constraints by having a large scalar-field 
mass in a region of high curvature.
Matter density perturbations in these models are enhanced
at small redshifts because of the presence of a coupling $Q$ 
that characterizes the strength between dark energy and non-relativistic matter.
We compute a convergence power spectrum of weak lensing 
numerically and show that the spectral index and the amplitude 
of the spectrum in the linear regime can be significantly modified 
compared to the $\Lambda$CDM model for large values of 
$|Q|$ of the order of unity. Thus weak lensing provides 
a powerful tool to constrain such large coupling scalar-tensor
models including $f(R)$ gravity.

\end{abstract}

\date{\today}

\maketitle

%%%%%%%%%%%%%%
\section{Introduction}
%%%%%%%%%%%%%%

The observations of the Supernovae Ia (SN Ia) in 1998 \cite{Perl} 
opened up a new research paradigm known as Dark Energy (DE). 
In spite of the tremendous effort over the past ten years,
we have not yet identified the origin of DE responsible for 
the late-time accelerated expansion.
Many DE models have been proposed so far to alleviate 
the theoretical problem of the cosmological constant 
scenario \cite{review,CST}.
We can broadly classify these models into two classes: 
(i) ``changing gravity'' models and (ii) ``changing matter'' models. 
The first class includes $f(R)$ gravity \cite{Capo}, 
scalar-tensor models \cite{stensor} and braneworld 
models \cite{brane}, whereas scalar-field models such as 
quintessence \cite{quin} and k-essence \cite{kes}
are categorized in the second class.

While changing matter models lead to dynamical evolution for 
the equation of state of DE, it is not easy to distinguish them from 
the cosmological constant scenario in current observations.
Meanwhile, if we change gravity from General Relativity, 
the models need to pass local gravity tests 
as well as cosmological constraints.
In this sense it is possible to place stringent 
experimental and observational constraints on
changing gravity models.

In fact there have been a burst of activities to search 
for viable modified gravity DE models.
In the so-called $f(R)$ gravity where $f$ is a function of 
the Ricci scalar $R$, it was found that the model
$f(R)=R-\alpha /R^n$ ($\alpha, n>0$) proposed in 
Refs.~\cite{Capo} is unable to satisfy 
the stability condition 
($f_{,RR} \equiv {\rm d}^2f/{\rm d}R^2>0$) 
for perturbations \cite{sta},
cosmological viability \cite{APT} and
local gravity constraints (LGC) \cite{lgc}.
Recently a number of authors proposed viable 
$f(R)$ DE models that satisfy all these 
requirements \cite{AGPT,Li,AT08,Hu,Star,Appleby,Tsuji,NO07,TUT}.
For example, the model $f(R)=R-\alpha R^n$ with 
$\alpha>0, 0<n<1$ is consistent with LGC 
for $n<10^{-10}$ \cite{CT} while 
at the same time satisfying stability and cosmological 
constraints. However it is difficult to distinguish 
this model from the $\Lambda$CDM cosmology 
because of the tight bound on the power $n$
coming from LGC.

The $f(R)$ models proposed by Hu and Sawicki \cite{Hu}
and Starobinsky \cite{Star} are designed to satisfy LGC 
in the region of high density where local gravity 
experiments are carried out.
Moreover it is possible to find an appreciable deviation 
from the $\Lambda$CDM model as the Universe
evolves from the matter-dominated epoch to the 
late-time accelerated era.
In fact the equation of state of DE in these models exhibits
peculiar evolution at small redshifts \cite{AT08,Tsuji}.
In addition, for the redshift smaller than a critical value 
$z_k$, the growth rate of matter density perturbations 
is larger than in the case of General Relativity \cite{Star,Tsuji}. 

Recently the analysis in $f(R)$ gravity was extended 
to a class of scalar-tensor DE models, i.e.,
Brans-Dicke theory with a scalar field 
potential $V(\phi)$ \cite{TUMTY}.
By introducing a constant $Q$ with the relation 
$1/(2Q^2)=3+2\omega_{\rm BD}$ ($\omega_{\rm BD}$ is 
a Brans-Dicke parameter), one can 
reduce this theory to the one given by the action 
(\ref{action}). The constant $Q$ characterizes 
the coupling between dark energy and non-relativistic
matter. If the scalar field $\phi$ is nearly massless,
the coupling is constrained to be $|Q| \lesssim 10^{-3}$
from solar system experiments \cite{TUMTY}.
However, if the field $\phi$ is massive in the region 
of high density, it is possible to satisfy LGC 
even when $|Q|$ is of the order of unity.
In fact, in the context of $f(R)$ gravity ($Q=-1/\sqrt{6}$), 
the models of Hu and Sawicki \cite{Hu} and Starobinsky \cite{Hu}
are designed in such a way that the field is sufficiently massive
in the regime $R \gg R_0$ ($R_0$ is the present cosmological 
Ricci scalar) and that the mass becomes lighter as 
$R$ approaches $R_0$. 
For general coupling $Q$, the potential given in Eq.~(\ref{po})
can be compatible with both local gravity 
and cosmological constraints.

The scalar-tensor models mentioned above show 
deviations from the $\Lambda$CDM model
at late times and hence they can leave 
a number of interesting observational signatures.
In Ref.~\cite{TUMTY} several bounds on the 
coupling $Q$ and model parameters were derived
by considering the evolution of matter density 
perturbations as well as LGC.
It was found that there exists allowed parameter
space of model parameters 
even when $|Q|$ is of the order of unity.

In this paper we shall study the effect of such modified
gravity models on weak lensing observations \cite{Bar}.
Since weak lensing carries the information of perturbations
at low redshifts, it is expected that this sheds light on
revealing the nature of 
DE \cite{Bena,Jain,Simpson,Heavens,Takada,AKS,Jaza,Vivi07,Bean}. 
In Refs.~\cite{Viviana,Uzan} a convergence power 
spectrum of weak lensing was derived in scalar-tensor theories
with the Lagrangian density 
${\cal L}=F(\phi)R/2-(\nabla \phi)^2/2-V(\phi)$.
In these theories a deflecting lensing potential $\Phi_{\rm wl}$
is modified compared to General Relativity due to the different 
evolution of gravitational potentials.
This gives rise to the change of 
the convergence power spectrum, which  
provides a powerful tool to distinguish 
modified gravity from the $\Lambda$CDM model.

The lensing potential $\Phi_{\rm wl}$ is sourced 
by matter density perturbations. 
The equation for matter perturbations
in scalar-tensor models was derived in 
Ref.~\cite{Boi} under the approximation on sub-horizon 
scales (see also Ref.~\cite{Hwang}). 
This analysis can be generalized to the theories with the 
Lagrangian density $f(R, \phi, X)$ 
(where $X=-(\nabla \phi)^2/2$), 
in which $\Phi_{\rm wl}$ was obtained 
analytically \cite{Tsuji07}.
The DGP braneworld model also leads to the modification 
to the lensing potential \cite{KM}. 
Thus the effect of modified gravity 
generally manifests itself in weak lensing observations.

In this work we focus on scalar-tensor models (\ref{action})
with a large coupling $Q$ and evaluate the convergence
power spectrum to find signatures of the modification of 
gravity in weak lensing. 
This analysis is general in the sense that $f(R)$ gravity
is included as a special case. 
In Sec.~\ref{mmodel} we review our scalar-tensor models
and present cosmological background equations to find dark 
energy dynamics. In Sec.~\ref{lensing} we derive the form of 
the convergence power spectrum as well as the equation for 
the deflecting potential $\Phi_{\rm wl}$.
In Sec.~\ref{signature} we compute the convergence 
spectrum numerically and estimate the effect of modified 
gravity on weak lensing. 
We conclude in Sec.~\ref{conclude}.

%%%%%%%%%%%%%%%%%%%%
\section{Modified gravity models}
\label{mmodel}
%%%%%%%%%%%%%%%%%%%%

We start with the following action 
\begin{eqnarray}
\label{action0}
S &=& \int {\rm d}^4 x\sqrt{-g} \left[ \frac12 \chi R
-\frac{\omega_{\rm BD}}{2\chi} (\nabla \chi)^2 -V(\chi) 
\right] \nonumber \\
& &+S_m (g_{\mu \nu}, \Psi_m)\,,
\end{eqnarray}
where $\chi$ is a scalar field coupled to the Ricci 
scalar $R$, 
$\omega_{\rm BD}$ is a constant parameter, $V(\chi)$
is a field potential, and $S_m$ is a matter action that depends on the
metric $g_{\mu \nu}$ and matter fields $\Psi_m$.
The action (\ref{action0}) corresponds to Brans-Dicke 
theory \cite{Brans} with a potential $V(\chi)$.
In the following we use the unit $8\pi G=1$, but we restore 
the bare gravitational constant $G$ when it is required.

Setting $\chi=F=e^{-2Q \phi}$, where $Q$ is a constant and 
$\phi=-1/(2Q)\,{\rm ln}\,\chi$ is a new scalar field, 
we find that the action (\ref{action0}) is equivalent to 
\begin{eqnarray}
\label{action}
S &=& \int {\rm d}^4 x\sqrt{-g} \left[ \frac12 F R
-\frac12 (1-6Q^2)F (\nabla \phi)^2-V \right]
\nonumber \\
& &+S_m (g_{\mu \nu}, \Psi_m)\,,
\end{eqnarray}
where $Q$ is related with the Brans-Dicke parameter $\omega_{\rm BD}$
via the relation $1/(2Q^2)=3+2\omega_{\rm BD}$ \cite{TUMTY}.
The $f(R)$ gravity corresponds to the coupling
$Q=-1/\sqrt{6}$, i.e., $\omega_{\rm BD}=0$ \cite{Chiba03}.

In the absence of the potential $V$ the Brans-Dicke parameter 
is constrained to be $\omega_{\rm BD}>4.0 
\times 10^4$ from 
solar system experiments \cite{lgccon}, 
which gives the bound $|Q|<2.5 \times 10^{-3}$.
If the potential $V$ is present, it is possible 
to satisfy solar system constraints even when $|Q|$ 
is of the order of unity by having a large mass 
in a high-curvature region.
In the context of $f(R)$ gravity, the following model is designed to 
satisfy LGC \cite{Tsuji}:
\begin{eqnarray}
\label{fR}
f(R) = R-\mu R_c [1-(R/R_c)^{-2n}]\,,
\end{eqnarray}
where $\mu$, $R_c$, $n$ are positive constants, and 
$R_c$ is roughly of the order of the present 
cosmological Ricci scalar $R_0$. 
Note that this satisfies the stability condition $f_{,RR}>0$
for $R \ge R_1$ ($R_1$ is a Ricci scalar at a late-time 
de-Sitter point) unlike the model 
$f(R)=R-\alpha /R^n$ ($\alpha, n>0$) \cite{Star,Tsuji}.
In the limit $R \gg R_c$ the above model approaches 
the $\Lambda$CDM model, which allows a possibility 
to be consistent with LGC in the region of high density.

In fact, the model (\ref{fR}) satisfies LGC 
for $n>0.9$ \cite{CT} through a chameleon mechanism \cite{KW} 
because of the presence of an effecive potential $V=(RF-f)/2$ with
the dynamical field 
$\phi=(\sqrt{6}/2)\,{\rm ln}\,F$.
The field potential in this case is given by 
\begin{eqnarray}
\label{pofR}
V(\phi)=\frac{\mu R_c}{2}
\left[ 1-\frac{2n+1}{(2n\mu)^{2n/(2n+1)}} 
(1-e^{2\phi/\sqrt{6}})^{\frac{2n}{2n+1}}\right].
\end{eqnarray}
The models proposed by Hu and Sawicki \cite{Hu}
and by Starobinsky \cite{Star} reduce to this form of the potential
in the high-curvature region ($R \gg R_c$) where local gravity 
experiments are carried out. 
When $R \gg R_c$ the field $\phi$ is almost frozen 
at instantaneous minima around $\phi=0$ 
characterized by the condition 
$e^{2\phi/\sqrt{6}}=1-2n\mu (R/R_c)^{-(2n+1)}$
with a large mass squared $M^2 \equiv V_{,\phi \phi} 
\propto \phi^{-\frac{2n+2}{2n+1}}$.
These minima are sustained by an effective coupling $Q$
between non-relativistic matter and the field $\phi$ \cite{CT}.

For arbitrary coupling $Q$ with the action (\ref{action}),
one can also construct viable models 
by generalizing the analysis of $f(R)$ gravity.
An explicit example of the potential consistent with LGC
is given by \cite{TUMTY}
\begin{eqnarray}
\label{po}
V(\phi)=V_1 \left[ 1-C (1-e^{-2Q\phi})^p \right]\,, 
\end{eqnarray}
where $V_1>0,~C>0,~0<p<1$.
This is motivated by the potential (\ref{pofR}), which means
that the $f(R)$ model (\ref{fR}) 
is recovered by setting $p=2n/(2n+1)$.
The analysis using the potential (\ref{po}) with the action 
(\ref{action}) is sufficiently general  
to understand essential features of modifed gravity
models that satisfy local gravity and cosmological constraints.
As $p$ gets closer to 1, the field mass in the region of 
high-curvature tends to be heavier so that the models are 
consistent with LGC.
In Ref.~\cite{TUMTY} it was found that the constraints 
coming from solar system tests and the violation of 
equivalence principle give the bounds 
$p>1-5/(9.6-{\rm ln}_{10}|Q|)$ and 
$p>1-5/(13.8-{\rm ln}_{10}|Q|)$, respectively.
In $f(R)$ gravity with the potential (\ref{pofR})
these bounds translate into the conditions $n>0.5$ and $n>0.9$, 
respectively \cite{CT}.

Let us review cosmological dynamics for the action (\ref{action}) 
with the potential (\ref{po})
in the flat Friedmann-Lemaitre-Robertson-Walker (FLRW) 
metric, ${\rm d}s^2=-{\rm d} t^2+a^2(t){\rm d}{\bf x}^2$, 
where $t$ is cosmic time and $a(t)$ is the scale factor.
As a source term for the matter action $S_m$, 
we take into account 
a non-relativistic fluid with energy density $\rho_{m}$
and a radiation with energy density $\rho_{\rm rad}$.
These obey the usual conservation equations
$\dot{\rho}_m+3H\rho_m=0$ and 
$\dot{\rho}_{\rm rad}+4H\rho_{\rm rad}=0$, 
where $H \equiv \dot{a}/a$.
The variation of the action (\ref{action}) leads to the 
following equations of motion:
\begin{eqnarray}
\label{be1}
& & \hspace*{-1.5em}
3FH^2=\frac12 (1-6Q^2) F\dot{\phi}^2+V
-3H\dot{F}+\rho_{m}+\rho_{\rm rad}\,, \\
\label{be2}
& & \hspace*{-1.5em}
2F\dot{H}=-(1-6Q^2)F \dot{\phi}^2
-\ddot{F}+H\dot{F}-\rho_{m}
-\frac43 \rho_{\rm rad}\,, \\
\label{be3}
& & \hspace*{-1.5em}
(1-6Q^2) F \left( \ddot{\phi}
+3H\dot{\phi}+\frac{\dot{F}}{2F}\dot{\phi}
\right)+V_{,\phi}+QFR=0\,,
\end{eqnarray}
where $R=6(2H^2+\dot{H})$.

In order to solve the background equations (\ref{be1})-(\ref{be3})
numerically, we introduce the dimensionless variables
\begin{eqnarray}
\label{x}
x_1=\frac{\dot{\phi}}{\sqrt{6}H}\,, \quad
x_2=\frac{1}{H}\sqrt{\frac{V}{3F}}\,, \quad
x_3=\frac{1}{H}\sqrt{\frac{\rho_{\rm rad}}{3F}}\,.
\end{eqnarray}
We also define 
\begin{eqnarray}
& &\Omega_{\rm DE} \equiv (1-6Q^2)x_1^2+x_2^2+
2\sqrt{6}Q x_1\,,\\
& &\Omega_{\rm rad} \equiv x_3^2\,,\\
& &\Omega_m \equiv
1-(1-6Q^2)x_1^2-x_2^2-2\sqrt{6}Q x_1-x_3^2\,,
\end{eqnarray}
which satisfy the relation $\Omega_{\rm DE}+
\Omega_{\rm rad}+\Omega_m=1$.
Using Eqs.~(\ref{be2}) and (\ref{be3}) we find
\begin{eqnarray}
\frac{\dot{H}}{H^2}&=& -\frac{1-6Q^2}{2} 
\biggl[ 3+3x_1^2-3x_2^2+x_3^2-6Q^2 x_1^2
\nonumber \\
& &~~~~~~~~~~~~~~
+2\sqrt{6}Q x_1 \biggr] 
+3Q (\lambda x_2^2-4Q),
\end{eqnarray}
where $\lambda=-V_{,\phi}/V$.
For the potential (\ref{po}) we have 
\begin{eqnarray}
\label{lambda}
\lambda=
\frac{2C\,p\,Q e^{-2Q \phi} (1-e^{-2Q\phi})^{p-1}}
{1-C(1-e^{-2Q\phi})^p}\,.
\end{eqnarray}
The effective equation of the system is defined by 
\begin{eqnarray}
w_{\rm eff} \equiv -1-2\dot{H}/(3H^2)\,.
\end{eqnarray}

Using Eqs.~(\ref{be1})-(\ref{be3}), we obtain 
the following equations
\begin{eqnarray}
\label{au1}
\frac{{\rm d} x_1}{{\rm d} N}
&=& \frac{\sqrt{6}}{2} (\lambda x_2^2-\sqrt{6} x_1)
+\frac{\sqrt{6}Q}{2} \biggl[ (5-6Q^2) x_1^2
\nonumber \\
& &+2 \sqrt{6}Q x_1-3x_2^2+x_3^2-1 \biggr]
-x_1 \frac{\dot{H}}{H^2}  \,, \\
\label{au2}
\frac{{\rm d} x_2}{{\rm d} N}
&=& \frac{\sqrt{6}}{2} (2Q-\lambda)x_1 x_2
-x_2 \frac{\dot{H}}{H^2} \,, \\
\label{au3}
\frac{{\rm d} x_3}{{\rm d} N}
&=& \sqrt{6} Q x_1 x_3-2x_3
-x_3 \frac{\dot{H}}{H^2} \,,
\end{eqnarray}
where $N\equiv{\rm ln}\,(a)$ is the number of e-foldings.
We note that the variable $F$ satisfies the equation of motion:
${\rm d}F/{\rm d}N=-2\sqrt{6}Q x_1 F$.

There exists a radiation fixed point: $(x_1,x_2,x_3)=(0,0,1)$
for this system. During radiation and matter eras, the field $\phi$ 
is stuck around the ``instantaneous'' minima characterized 
by the condition $V_{,\phi}+QFR=0$, i.e., 
\begin{eqnarray}
\label{Qphi}
2Q\phi_m \simeq \left( \frac{2V_1pC}{\rho_m} 
\right)^{\frac{1}{1-p}} \ll 1\,,
\end{eqnarray}
where we used the fact that $V_1$ is of the order of the squared 
of the present Hubble parameter $H_0$ so that 
the potential (\ref{po}) is responsible for 
the accelerated expansion today.
Note that we have $F=e^{-2Q\phi_m} \simeq 1$
under the condition (\ref{Qphi}).
In this region the quantity $|\lambda|$ defined in Eq.~(\ref{lambda}) 
is much larger than unity. 
The field value $|\phi_m|$ increases as the system enters 
the epoch of an accelerated expansion, 
which leads to the decrease of $|\lambda|$.
The matter-dominated epoch is realized by the instantaneous fixed
point characterized by $(x_1,x_2,x_3)=(\sqrt{6}/(2\lambda),
[(3+2Q\lambda-6Q^2)/2\lambda^2]^{1/2},0)$
with $\Omega_m=1-(3-12Q^2+7Q\lambda)/\lambda^2 \simeq 1$
and $w_{\rm eff}=-2Q/\lambda \simeq 0$ 
(because $|\lambda| \gg 1$ in this regime).
In the presence of the coupling $Q$ there exists a de-Sitter point 
characterized by $(x_1,x_2,x_3)=(0,1,0)$, 
$\Omega_m=0$ and $w_{\rm eff}=-1$, which corresponds to 
$\lambda=4Q$. This solution is stable for 
${\rm d}\lambda/{\rm d}\phi<0$ \cite{TUMTY}
and hence can be used for the late-time accelerated expansion.
See Ref.~\cite{TUMTY} for detailed analysis about  
the background cosmological evolution.

The mass squared, $M^2=V_{,\phi \phi}$, is given by 
\begin{eqnarray}
\label{Msq}
M^2=4V_1 Cp Q^2(1-pe^{-2Q\phi})
(1-e^{-2Q\phi})^{p-2} e^{-2Q\phi}.
\end{eqnarray}
Plugging the field value $\phi_m$ into Eq.~(\ref{Msq}), 
we find
\begin{eqnarray}
\label{Msq2}
M^2 \simeq \frac{1-p}{(2^p p C)^{1/(1-p)}}
Q^2 \left( \frac{\rho_m}{V_1} 
\right)^{\frac{2-p}{1-p}}V_1\,.
\end{eqnarray}

Since the energy density $\rho_m$ is much larger than 
$V_1$ during the radiation and matter eras, 
we have that $M^2 \gg V_1 \sim H_0^2$.
The mass squared $M^2$ decreases to the order of
$V_1$ after the system enters the accelerated epoch.
This evolution of the field mass leads to an interesting 
observational signature in weak lensing observations, 
as we will see in subsequent sections.

%%%%%%%%%%%%%%%%%%%%%%%%%%%%%%%%%%%%%%%
\section{Weak lensing}
\label{lensing}
%%%%%%%%%%%%%%%%%%%%%%%%%%%%%%%%%%%%%%%

Let us consider a perturbed metric about the 
flat FLRW background in the longitudinal gauge:
\begin{eqnarray}
\rd s^2=-(1+2\Phi) \rd t^2
+a^2(t) (1-2\Psi) \delta_{ij}
\rd x^i \rd x^j\,,
\end{eqnarray}
where scalar metric perturbations $\Phi$ and $\Psi$ do not 
coincide with each other in the absence of an anisotropic stress.
Matter density perturbations $\delta_m$ in the pressureless matter contribute to the source term for the gravitational potentials
$\Phi$ and $\Psi$.
The equation of $\delta_m$ for the action (\ref{action}) 
was derived in Ref.~\cite{TUMTY} under an approximation 
on sub-horizon scales \cite{Boi,CST,Tsuji07}.
Provided that the oscillating mode of the field perturbation $\delta \phi$
does not dominate over the matter-induced mode
at the initial stage of the matter era, 
we obtain the following approximate equation 
\begin{eqnarray}
\label{delmap}
\ddot{\delta}_m+2H\dot{\delta}_m-4\pi G_{\rm eff}
\rho_m \delta_m  \simeq 0\,,
\end{eqnarray}
where the effective gravitational ``constant'' is given by
\begin{eqnarray}
\label{Geff}
G_{\rm eff}=\frac{1}{8\pi F}
\frac{(k^2/a^2)(1+2Q^2)F+M^2}
{(k^2/a^2)F+M^2}\,.
\end{eqnarray}
Here $k$ is a comoving wavenumber and $M^2$
is given in Eq.~(\ref{Msq}) for the potential (\ref{po}).
Using the derivative with respect to $N$, Eq.~(\ref{delmap}) can be written as 
\begin{eqnarray}
\label{delmap2}
& & \frac{{\rm d}^2 \delta_m}{{\rm d}N^2}+
\left( \frac12 - \frac32 w_{\rm eff} \right)
\frac{{\rm d} \delta_m}{{\rm d}N} \nonumber \\
& & -\frac32 \Omega_m
\frac{(k^2/a^2)(1+2Q^2)F+M^2}
{(k^2/a^2)F+M^2}\delta_m \simeq 0\,.
\end{eqnarray}
The gravitational potentials $\Phi$ and $\Psi$ satisfy 
\begin{eqnarray}
\label{Phi}
& &\frac{k^2}{a^2} \Phi \simeq
-\frac{\rho_m}{2F}
\frac{(k^2/a^2)(1+2Q^2)F+M^2}
{(k^2/a^2)F+M^2} \delta_m\,,\\
\label{Psi}
& &\frac{k^2}{a^2} \Psi \simeq
-\frac{\rho_m}{2F}
\frac{(k^2/a^2)(1-2Q^2)F+M^2}
{(k^2/a^2)F+M^2} \delta_m\,. 
\end{eqnarray}

In order to confront our model with weak lensing
observations, we define the so-called
deflecting potential \cite{Uzan}
\begin{eqnarray}
\Phi_{\rm wl} \equiv \Phi+\Psi\,,
\end{eqnarray}
together with the effective density field 
\begin{eqnarray}
\label{del}
\delta_{\rm eff} \equiv 
-\frac{a}{3H_0^2 \Omega_{m,0}}
k^2\,\Phi_{\rm wl}\,,
\end{eqnarray}
where the subscript ``0'' represents the present 
values and we set $a_0=1$.
Using the relation 
\begin{eqnarray}
\label{rhom}
\rho_m=3F_0H_0^2 \Omega_{m,0}/a^3\,,
\end{eqnarray}
together with Eqs.~(\ref{Phi}) and (\ref{Psi}), 
we get 
\begin{eqnarray}
\label{Phiwl}
\Phi_{\rm wl}=-\frac{a^2}{k^2}
\frac{\rho_m}{F} \delta_m\,,\quad
\delta_{\rm eff}=\frac{F_0}{F}\delta_m\,.
\end{eqnarray}

We write the angular position of a source to be $\vec{\theta}_S$
and the direction of weak lensing observation to be $\vec{\theta}_I$.
The deformation of the shape of galaxies is characterized
by the amplification matrix ${\cal A}=\rd \vec{\theta}_S/
\rd \vec{\theta}_I$.
The components of ${\cal A}$ are given by \cite{Bar,Uzan}
\begin{eqnarray}
\label{A}
{\cal A}_{\mu \nu}=I_{\mu \nu}
-\int_0^{\chi} \frac{\chi'(\chi-\chi')}{\chi}
\partial_{\mu \nu} \Phi_{\rm wl} 
[\chi' \vec{\theta}, \chi'] {\rm d} \chi'\,,
\end{eqnarray}
where $\chi$ is the comoving radial distance satisfying 
the relation $\rd \chi=-\rd t/a(t)$ along the geodesic.
In terms of the redshift defined by $z=1/a-1$, 
we have that  
\begin{eqnarray}
\label{chiz}
\chi(z)=\int_0^z \frac{{\rm d}z'}{H(z')}\,.
\end{eqnarray}
The convergence $\kappa$ and the shear $\vec{\gamma}
=(\gamma_1,\gamma_2)$
can be derived from the components of the 
$2 \times 2$ matrix ${\cal A}$, as
\begin{eqnarray}
\label{kappa}
\kappa=1-\frac12 {\rm Tr}\,{\cal A}\,,\quad
\vec{\gamma}=\left([{\cal A}_{22}-{\cal A}_{11}]/2, 
{\cal A}_{12} \right)\,.
\end{eqnarray}
If we consider a redshift distribution $p(\chi){\rm d}\chi$
of the source, the convergence is given by  
$\kappa (\vec{\theta})=\int p(\chi) 
\kappa (\vec{\theta},\chi){\rm d}\chi$.
Using Eqs.~(\ref{del}), (\ref{A}) and (\ref{kappa}) we obtain 
\begin{eqnarray}
\kappa (\vec{\theta})=\frac32 H_0^2 \Omega_{m,0}
\int_0^{\chi_H} g(\chi)\chi 
\frac{\delta_{\rm eff}[\chi\,\vec{\theta},\chi]}{a} \rd \chi\,,
\end{eqnarray}
where $\chi_H$ is the maximum distance to the source and 
\begin{eqnarray}
g(\chi) \equiv \int_{\chi}^{\chi_H}
p(\chi') \frac{\chi'-\chi}{\chi'}
\rd \chi'\,.
\end{eqnarray}

Since the convergence is a function on the 2-sphere it can 
be expanded in the form $\kappa (\vec{\theta})=
\int \hat{\kappa} (\vec{\ell}) e^{i \vec{\ell}\cdot \vec{\theta}}
\frac{{\rm d}^2 \vec{\ell}}{2\pi}$, where
$\vec{\ell}=(\ell_1, \ell_2)$ with 
$\ell_1$ and $\ell_2$ integers.
Defining the power spectrum of the shear to be 
$\langle \hat{\kappa} (\vec{\ell}) \hat{\kappa}^* 
(\vec{\ell'})\rangle
=P_{\kappa}(\vec{\ell}) \delta^{(2)} (\vec{\ell}-\vec{\ell'})$, 
one can show that the convergence 
has a same power spectrum as $P_{\kappa}$ \cite{Bar}. 
It is given by \cite{Uzan}
\begin{eqnarray}
P_{\kappa} (\ell)=\frac{9H_0^4 \Omega_{m,0}^2}{4}
\int_0^{\chi_H} \left[ \frac{g(\chi)}{a(\chi)} \right]^2
P_{\delta_{\rm eff}} \left[ \frac{\ell}{\chi}, \chi 
\right] \rd \chi\,.
\end{eqnarray}

In our scalar-tensor theory we have $P_{\delta_{\rm eff}}=
(F_0/F)^2 P_{\delta_m}$ from Eq.~(\ref{Phiwl}), where
$P_{\delta_m}$ is the matter power spectrum.
In the following we assume that the sources are located 
at the distance $\chi_s$ (corresponding to the redshift $z_s$), 
which then gives $p(\chi)=\delta (\chi-\chi_s)$ and 
$g(\chi)=(\chi_s-\chi)/\chi_s$.
This leads to the following convergence spectrum
\begin{equation}
\label{Pkappa}
\hspace*{-0.3em}
P_{\kappa} (\ell)=\frac{9H_0^4 \Omega_{m,0}^2}{4}
\int_0^{\chi_s} \left( \frac{\chi_s-\chi}{\chi_s a} 
\frac{F_0}{F} \right)^2
P_{\delta_m} \left[ \frac{\ell}{\chi}, \chi 
\right] \rd \chi. 
\end{equation}

Let us consider the action (\ref{action}) with the 
potential (\ref{po}).
In the deep matter era where the Ricci scalar $R$ is 
much larger than $H_0^2$, we have $M^2/F \gg k^2/a^2$
and $F \simeq 1$ for the wavenumber $k$ 
relevant to the matter power spectrum \cite{TUMTY}. 
Since $G_{\rm eff} \simeq G$
in this regime from Eq.~(\ref{Geff}), 
the perturbations evolve in a standard way:
$\delta_m \propto t^{2/3}$ and 
$\Phi_{\rm wl}={\rm constant}$.
Meanwhile, at the late epoch of the matter era,  
the system can enter a stage characterized
by the condition $M^2/F \ll k^2/a^2$.
Since $G_{\rm eff} \simeq (1+2Q^2)/8\pi F$ 
during this stage, the perturbations 
evolve in a non-standard way:
\begin{equation}
\label{growth}
\delta_m \propto t^{(\sqrt{25+48Q^2}-1)/6}\,,\quad
\Phi_{\rm wl} \propto t^{(\sqrt{25+48Q^2}-5)/6}.
\end{equation}
The critical redshift $z_k$ at $M^2/F = k^2/a^2$ 
can be estimated as 
\begin{equation}
\label{zk}
z_k \simeq \left[\left( \frac{k^2}{H_0^2}
\frac{1}{Q^2(1-p)} \right)^{1-p} \frac{2^p pC}
{(3F_0 \Omega_{m,0})^{2-p}}
\frac{V_1}{H_0^2} \right]^{\frac{1}{4-p}}-1.
\end{equation}
As long as $z_k \gtrsim 1$ it is expected that the effect of modified 
gravity manifests itself in weak lensing observations.

Since the evolution of perturbations is similar to that in the 
$\Lambda$CDM model at an early epoch characterized 
by the condition $z \gg z_k$, the deflecting potential 
$\Phi_{\rm wl}$ at late times is given by \cite{Dodelson}
\begin{eqnarray}
\label{Phikdef}
\Phi_{\rm wl} (k, a)=\frac{9}{10} \Phi_{\rm wl}(k, a_i)
T(k) \frac{D(k, a)}{a}\,,
\end{eqnarray}
where $\Phi_{\rm wl}(k, a_i) \simeq 2\Phi (k, a_i)$ corresponds
to the initial deflecting potential generated during inflation, 
$T(k)$ is a transfer function that describes the epochs of 
horizon crossing and radiation/matter 
transition ($50 \lesssim z \lesssim 10^6$), 
and $D(k,a)$ is the growth function at late times defined by 
$D(k,a)/a=\Phi_{\rm wl}(a)/\Phi_{\rm wl} (a_I)$
($a_I$ corresponds to the scale factor 
at a redshift $1 \ll z_I<50$).

Since we are interested in the case where 
the transition redshift $z_k$ is smaller than 50, 
we can use the standard transfer 
function of Bardeen {\it et al.} \cite{BBKS}:
\begin{eqnarray}
T(x) &=& 
\frac{{\rm ln} (1+0.171x)}{0.171 x}
\biggl [1.0+0.284x+(1.18x)^2 \nonumber \\
& &+(0.399x)^3 
+(0.490x)^4 \biggr]^{-0.25},
\end{eqnarray}
where $x \equiv k/k_{\rm EQ}$ and 
$k_{\rm EQ}=0.073\,\Omega_{m,0}h^2$\,Mpc$^{-1}$.

In the $\Lambda$CDM model the growth function during the matter-dominated 
epoch ($\Omega_m=1$) is scale-independent: 
$D(k, a)=a$.\footnote{Note 
that in the late-time accelerated epoch the growth 
of matter pertubations is no longer described by $D(a)=a$.}
In our scalar-tensor model the mass squared $M^2$ given 
in Eq.~(\ref{Msq2}) evolves as 
$M^2 \propto t^{-2(2-p)/(1-p)}$, 
which implies that the transition time $t_k$
at $M^2/F=k^2/a^2$ has a scale-dependence 
$t_k \propto k^{-\frac{3(1-p)}{4-p}}$ \cite{TUMTY}.
This leads to the scale-dependent growth 
of metric perturbations.

Using Eqs.~(\ref{del}) and (\ref{Phikdef}) we obtain 
the matter perturbation $\delta_m$ at the 
redshift $z<z_I$:
\begin{eqnarray}
\delta_m (k, a)=-\frac{3}{10} \frac{F}{F_0}
\frac{k^2}{\Omega_{m,0}H_0^2}
\Phi_{\rm wl} (k, a_i) T(k) D(k, a)\,.
\end{eqnarray}
The initial power spectrum generated 
during inflation is $P_{\Phi_{\rm wl}} \equiv 
4|\Phi|^2 =(200 \pi^2/9k^3)(k/H_0)^{n_s-1} \delta_H^2$, 
where $n_s$ is the spectral index and $\delta_H^2$ 
is the amplitude of $\Phi_{\rm wl}$. 
Then the power spectrum, 
$P_{\delta_m} \equiv |\delta_m|^2$, is given by 
\begin{equation}
\label{delm}
P_{\delta_m}(k,a)=
2\pi^2 \left(\frac{F}{F_0}\right)^2 
\frac{k^{n_s}}{\Omega_{m,0}^2 H_0^{n_s+3}}
\delta_H^2 T^2(k) D^2(k, a).
\end{equation}

{}From Eqs.~(\ref{Pkappa}) and (\ref{delm}) we get 
\begin{eqnarray}
\label{Pkappa2}
P_{\kappa} (\ell) &=&
\frac{9\pi^2}{2}
\int_0^{z_s} \left( 1-\frac{X}{X_s} \right)^2
\frac{1}{E(z)} \delta_H^2 \nonumber \\
& & \times 
\left( \frac{\ell}{X} \right)^{n_s} T^2(x)
\left(\frac{\Phi_{\rm wl}(z)}
{\Phi_{{\rm wl}}(z_I)} \right)^2 \rd z\,,
\end{eqnarray}
where 
\begin{eqnarray}
E(z)=\frac{H(z)}{H_0}\,,\quad 
X=H_0 \chi\,,\quad
x=\frac{H_0}{k_{\rm EQ}}
\frac{\ell}{X}\,.
\end{eqnarray}
{}From Eq.~(\ref{chiz}) the quantity $X$ satisfies the 
differential equation ${\rd X}/{\rd z}=1/E(z)$.
In the following we use the 
value $z_s=1$ in our numerical simulations.

%%%%%%%%%%%%%%%%%%%%%%%%%%%%%%%%%%%%%%%
\section{Observational signatures of modified gravity}
\label{signature}
%%%%%%%%%%%%%%%%%%%%%%%%%%%%%%%%%%%%%%%

When $Q \neq 0$ the evolution of $\delta_m$
during the time-interval $t_k<t<t_\Lambda$ 
(where $t_\Lambda$ is the time
at $\ddot{a}=0$) is given by Eq.~(\ref{growth}), 
whereas $\delta_m \propto t^{2/3}$ in 
the $\Lambda$CDM model ($Q=0$).
Hence, at time $t_{\Lambda}$, the power
spectrum for $Q \neq 0$ exhibits a difference compared
to the $\Lambda$CDM model \cite{TUMTY}: 
\begin{eqnarray}
\frac{P_{\delta_m}(t_\Lambda)}
{P_{\delta_m}^{\Lambda {\rm CDM}} (t_\Lambda)}
=\left(\frac{t_{\Lambda}}
{t_k}\right)^{2\left( \frac{\sqrt{25+48Q^2}-1}
{6}-\frac23  \right)}
\propto k^{\Delta n (t_\Lambda)}\,,
\end{eqnarray}
where 
\begin{eqnarray}
\label{deles}
\Delta n (t_\Lambda)=\frac{(1-p)
(\sqrt{25+48Q^2}-5)}{4-p}\,.
\end{eqnarray}
In order to derive the difference $\Delta n (t_0)$ at the
present epoch, we need to solve perturbation equations 
numerically by the time $t_0$.
However, as long as $z_k$ is larger than the order of unity, 
the growth rate of $\delta_m$ during the  time-interval
$t_{\Lambda}<t<t_0$ hardly depends on $k$ for fixed $Q$.
Hence it is expected that the analytic estimation (\ref{deles})
does not differ much from $\Delta n (t_0)$
provided $z_k \gg 1$.

We start integrating the background equations 
(\ref{au1})-(\ref{au3}) from the deep matter era 
and identify the present epoch by the condition 
$\Omega_m=0.28$. We then run the code again from 
$z=z_I\,(<50)$ to $z=0$
in order to solve the perturbation 
equations (\ref{delmap2}) and (\ref{Phiwl}).
Since we are considering the case in which $z_k$ is 
smaller than $z_I$, the initial conditions for matter 
perturbations are chosen to be 
$\frac{{\rm d}\delta_m}{{\rm d}N}=\delta_m$ 
(i.e., those for the $\Lambda$CDM model).

\begin{figure}
\includegraphics[height=3.0in,width=3.4in]{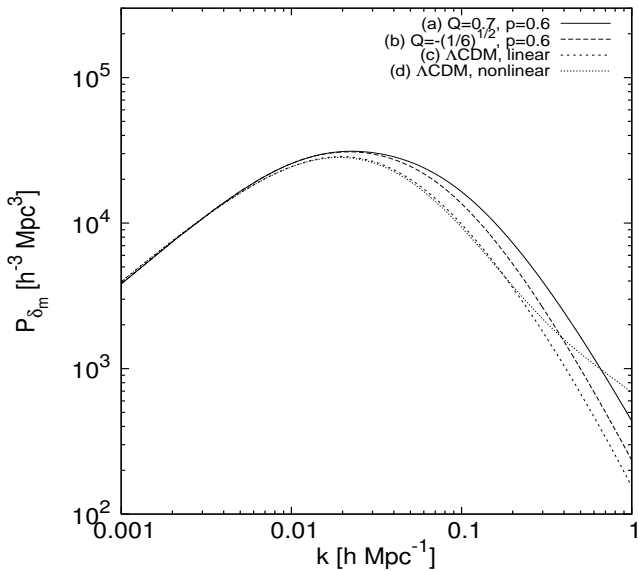}
\caption{\label{fig1} The matter power spectra $P_{\delta_m}(k)$
at the present epoch for (a) $Q=0.7$, $p=0.6$, $C=0.9$, 
(b) $Q=-1/\sqrt{6}$, $p=0.6$, $C=0.9$,
(c) the $\Lambda$CDM model, and
(d) the $\Lambda$CDM model with a nonlinear halo-fitting
($\sigma_8=0.78$ and shape parameter $\Gamma=0.2$).
The model parameters are $\Omega_{m,0}=0.28$, 
$H_0=3.34 \times 10^{-4}\,h$\,Mpc$^{-1}$, 
$n_s=1$ and $\delta_H^2=3.2 \times 10^{-10}$.
In the cases (a) and (b) we start integrating Eqs.~(\ref{au1})-(\ref{au3})
with initial conditions $(x_1,x_2,x_3)=
(0,[(3+2Q\lambda-6Q^2)/2\lambda^2]^{1/2},0)$
and $F-1=-10^{-8}$. }
\end{figure}

In Fig.~\ref{fig1} we plot the matter power spectra 
at the present epoch for (a) $Q=0.7$, $p=0.6$, $C=0.9$, 
(b) $Q=-1/\sqrt{6}$, $p=0.6$, $C=0.9$, 
(c) the $\Lambda$CDM model, and 
(d) the $\Lambda$CDM model with a nonlinear 
halo-fitting \cite{halo}.
Since we do not take into account nonlinear 
effects in the cases (a)-(c), these results are trustable in the linear 
regime $k \lesssim 0.2h$\,Mpc$^{-1}$.

In the case (b), which corresponds to $f(R)$ gravity
with $n=0.75$ in the model (\ref{fR}),
the spectrum shows a deviation from 
the $\Lambda$CDM model for $k>0.01h$\,Mpc$^{-1}$.
On the scales $k=0.01h$\,Mpc$^{-1}$ and 
$k=0.1h$\,Mpc$^{-1}$ the critical redshifts at 
$M^2/F=k^2/a^2$  are given by 
$z_k=2.995$ and $z_k=5.868$, respectively. 
Numerically we find $\Delta n (t_0)=0.017$ and 
$\Delta n(t_0)=0.119$ for $k=0.01h$\,Mpc$^{-1}$ 
and $k=0.1h$\,Mpc$^{-1}$ respectively, 
whereas the estimation (\ref{deles}) gives the value 
$\Delta n(t_{\Lambda})=0.088$.
Since $z_k$ decreases for smaller $k$, 
the analytic estimation (\ref{growth}) obtained by using 
the condition $z_k \gg 1$ tends to be invalid on larger scales.
This is the main reason of the discrepancy between 
$\Delta n (t_0)$ and $\Delta  n(t_\Lambda)$ found for 
$k<0.1h$\,Mpc$^{-1}$.
We checked that $\Delta n (t_0)$ approaches
the analytic value $\Delta n (t_\Lambda)=0.088$ 
on smaller scales, e.g., $\Delta n (t_0)=0.089$ 
for $k=4.3h$\,Mpc$^{-1}$.

For larger $|Q|$ the growth rate of $\delta_m$
increases in the regime $z_{\Lambda}<z<z_k$, 
which alters the shape of the matter power spectrum. 
In the case (a) of Fig.~\ref{fig1} we numerically find that 
$\Delta n (t_0)=0.323$ on the scale $k=0.1h$\,Mpc$^{-1}$, 
while the estimation (\ref{deles}) gives
$\Delta n (t_\Lambda)=0.231$.
Again this analytic estimation is in a better agreement with 
$\Delta n (t_0)$ on smaller scales, e.g.,
$\Delta n (t_0)=0.244$ for $k=4.3h$\,Mpc$^{-1}$.
In Fig.~\ref{fig1} we also show
the matter power spectrum in the $\Lambda$CDM model 
derived by using the nonlinear halo-fit \cite{halo}.
This gives rise to an enhancement of the power
in the nonlinear regime ($k>0.2h$\,Mpc$^{-1}$).
The spectrum in the case (a) exhibits a significant 
difference compared to this halo-fit $\Lambda$CDM
spectrum even for $k<0.2h$\,Mpc$^{-1}$, which implies that 
our linear analysis is enough to place stringent 
constraints on model parameters $Q$ and $p$ 
from observations of galaxy clustering.

\begin{figure}
\includegraphics[height=3.0in,width=3.5in]{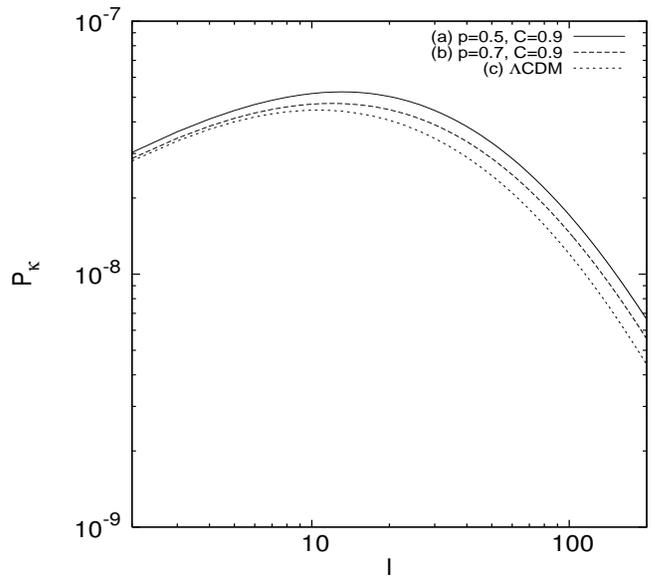}
\caption{\label{fig2} The convergence power spectrum $P_{\kappa}(\ell)$
in $f(R)$ gravity ($Q=-1/\sqrt{6}$) for the cases: 
(a) $p=0.5$, $C=0.9$ and (b) $p=0.7$, $C=0.9$. 
We also show the spectrum in the $\Lambda$CDM model. 
Other model parameters are chosen 
similarly as in the case of Fig.~\ref{fig1}.
}
\end{figure}
\begin{figure}
\includegraphics[height=3.0in,width=3.5in]{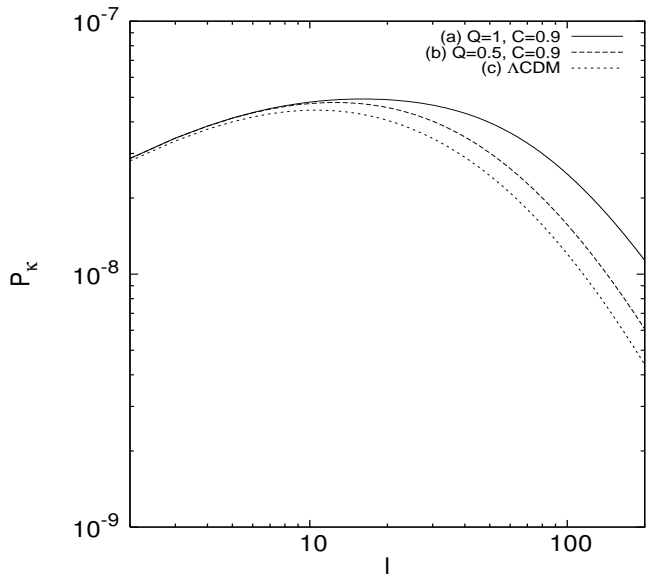}
\caption{\label{fig3} The convergence power spectrum 
$P_{\kappa}(\ell)$ for $p=0.7$ with two cases:  
(a) $Q=1$, $C=0.9$ and (b) $Q=0.5$, $C=0.9$
together with the $\Lambda$CDM spectrum.
Other model parameters are chosen 
similarly as in the case of Fig.~\ref{fig1}.
}
\end{figure}

Let us next proceed to the convergence power spectrum 
of weak lensing.
Compared to the matter power spectrum
the wavenumber $k$ is replaced by $k=\ell/\chi$.
In the deep matter era the evolution of the Hubble parameter 
can be approximated as $H^2(z) \simeq H_0^2 \Omega_{m,0} (1+z)^3$, 
which gives $\chi \simeq 2/(H_0 \Omega_{m,0}^{1/2})
={\rm constant}$.
Hence the time $t_{\ell}$ at 
$M^2/F=(\ell/\chi)^2/a^2$ has an
$\ell$-dependence $t_{\ell} \propto 
\ell^{-\frac{3(1-p)}{4-p}}$, provided 
this transition occurs at the redshift $z_{\ell} \gg 1$.

Since $\Phi_{\rm wl} \simeq {\rm constant}$  
for $t_I<t<t_{\ell}$ and 
$\Phi_{\rm wl} \propto t^{(\sqrt{25+48Q^2}-5)/6}$
for $t_{\ell}<t<t_{\Lambda}$, we have that  
\begin{eqnarray}
\frac{\Phi_{\rm wl}(z_{\Lambda})}
{\Phi_{\rm wl}(z_I)} \simeq
\left( \frac{t_\Lambda}{t_{\ell}} \right)
^{(\sqrt{25+48Q^2}-5)/6}\,.
\end{eqnarray}
As long as $z_{\ell} \gg 1$, the evolution of 
$\Phi_{\rm wl}$ during the time-interval 
$t_{\Lambda}<t<t_0$ is almost independent 
of $\ell$ for a fixed value of $Q$.
Then we obtain the following $\ell$-dependence
for $0<z<z_{\Lambda} \sim z_s$:
\begin{eqnarray}
\left( \frac{\Phi_{\rm wl}(z)}
{\Phi_{\rm wl}(z_I)} \right)^2 
\propto \ell^{\frac{(1-p)
(\sqrt{25+48Q^2}-5)}{4-p}}\,.
\end{eqnarray}
{}From Eq.~(\ref{Pkappa2}) this leads to a difference 
of the spectral index of the convergence spectrum
compared to the $\Lambda$CDM model:
\begin{eqnarray}
\label{Pkaell}
\frac{P_{\kappa} (\ell)}
{P_{\kappa}^{\Lambda {\rm CDM}} (\ell)}
\propto
\ell^{\Delta n}\,,
\end{eqnarray}
where $\Delta n$ is the same as $\Delta n (t_\Lambda)$
given in Eq.~(\ref{deles}).
We caution again that the estimation (\ref{Pkaell})  
is valid for $z_{\ell} \gg 1$.

In Fig.~\ref{fig2} we plot the convergence spectrum
in $f(R)$ gravity for two different values of $p$
together with the $\Lambda$CDM spectrum.
We focus on the linear regime characterized by 
$\ell \lesssim 200$.
Since the $\Lambda$CDM model corresponds to 
the limit $n \to \infty$ in Eq.~(\ref{fR}), the power 
$p=2n/(2n+1)$ approaches 1 in this limit.
The deviation from the $\Lambda$CDM model 
becomes important for smaller $p$ away from 1.

When $p=0.7$, for example, Fig.~\ref{fig2}
shows that such a deviation becomes significant 
for $\ell \gtrsim 10$.
Numerically we get $\Delta n=0.056$ at $\ell=200$, which
is slightly smaller than the analytic value 
$\Delta n=0.068$ estimated by Eq.~(\ref{Pkaell}). 
The main reason for this 
difference is that the critical redshift 
$z_{\ell}=3.258$ at $\ell=200$
is not very much larger than unity.

When $p=0.5$ the deflecting potential 
$\Phi_{\rm wl}$ is amplified even for small $\ell$ 
($\lesssim 10$), which is associated with the fact that 
$z_{\ell}$ is greater than 1 even for $\ell>2$.
For example we find that $z_{\ell}=1.386$ for $\ell=5$.
In this case the system enters the non-standard 
regime ($z<z_\ell$) before entering the epoch 
of an accelerated expansion ($z<z_{\Lambda} \sim 1$), 
which leads to the amplification of $\Phi_{\rm wl}$.
This changes the total amplitude of $P_{\kappa} (\ell)$
relative to the $\Lambda$CDM model.
The numerical value of $\Delta n$ at $\ell=200$ 
is found to be $\Delta n=0.084$ for $p=0.5$.
Since $\Delta n$ increases for smaller $p$, this information 
is useful to place a lower bound on $p$ in $f(R)$ gravity from 
weak lensing observations.

In Fig.~\ref{fig3} the convergence spectrum 
for $p=0.7$ is plotted for two different values 
of $Q$ together with the $\Lambda$CDM spectrum.
We note that the transition redshift $z_{\ell}$ 
decreases for larger $|Q|$, see Eq.~(\ref{zk}).
Hence the deviation from the $\Lambda$CDM model
is insignificant for small $\ell$, unless we choose smaller
values of $p$.
However the spectrum is strongly modified for 
$\ell \gtrsim 10$ with the increase of $|Q|$.
The numerical values of $\Delta n$ at $\ell=200$ are
found to be $\Delta n=0.084$ and $\Delta n=0.311$ for 
$Q=0.5$ and $Q=1$, respectively.
Hence it should be possible to derive an upper bound on
the strength of the coupling $Q$ by using observational 
data of weak lensing. 

%%%%%%%%%%%%%%%%%%%%%%%%%%%%%%%%%%%%%%%
\section{Conclusions}
\label{conclude}
%%%%%%%%%%%%%%%%%%%%%%%%%%%%%%%%%%%%%%%

We have discussed the signature of modified gravity in weak lensing 
observations. Our model is described by the action 
(\ref{action}) with a constant coupling $Q$, which 
is equivalent to Brans-Dicke theory with a field potential $V$.
This theory includes $f(R)$ gravity as a special case ($Q=-1/\sqrt{6}$).
The scalar-field potential $V(\phi)$ can be designed 
to satisfy local gravity constraints through a chameleon 
mechanism. The representative potential that satisfies LGC 
is given in Eq.~(\ref{po}), which is motivated by viable $f(R)$
models proposed by Hu and Sawicki \cite{Hu} and 
by Starobinsky \cite{Star}.
Note that most of past works in scalar-tensor dark energy 
models restricted the analysis in the small coupling region  
($|Q| \lesssim 10^{-3}$).
In this paper we focused on the large $|Q|$ region 
in which a significant difference from 
the $\Lambda$CDM model can be expected
in weak lensing observations.

Cosmologically these models can show deviations from the  
$\Lambda$CDM model at late epochs of the 
matter-dominated era. The growth rate of 
matter density perturbations gets larger 
for redshifts smaller than a critical value $z_k$.
Since $z_k$ increases for larger $k$, the matter 
power spectrum is subject to change on smaller scales.
We evaluated the matter power spectrum $P_{\delta_m}(k)$
numerically and showed that the spectral index and the amplitude
of $P_{\delta_m}(k)$ can be significantly modified for larger
values of $|Q|$.

The non-standard evolution of matter perturbations affects
the convergence power spectrum $P_{\kappa}(\ell)$ of weak lensing.
As long as the transition redshift $z_{\ell}$ is larger than 
the order of unity, one can estimate the difference $\Delta n$ 
of spectral indices between modified gravity and the $\Lambda$CDM
cosmology to be $\Delta n \simeq (1-p)
(\sqrt{25+48Q^2}-5)/(4-p)$ with $0<p<1$.
In $f(R)$ gravity the parameter $n$ for the model (\ref{fR})  
is linked with the parameter $p$ via the relation $p=2n/(2n+1)$.
The limit $p \to 1$ (i.e., $n \to \infty$) corresponds to 
the $\Lambda$CDM model, in which case we have $\Delta n \to 0$.
The difference of the convergence spectrum relative to 
the $\Lambda$CDM case is significant for $p$ away from 1.
As seen in Fig.~\ref{fig2} (which corresponds to the case
$Q=-1/\sqrt{6}$), the spectral index and the amplitude
of $P_{\kappa}(\ell)$ are modified for smaller
values of $p$.

If we take larger values of $|Q|$,
the convergence spectrum deviates from that in 
the $\Lambda$CDM model
more significantly. This situation is clearly seen in the numerical 
simulation of Fig.~\ref{fig3}.  
It should be possible to place strong observational constraints
on the parameters $Q$ and $p$ by using observational 
data of weak lensing and the matter power spectrum, which 
we leave for future work.
We hope that some signatures of modified gravity 
can be detected in future high-precision observations
to reveal the origin of dark energy.

%%%%%%%%%%%%%%%%%%%%%%%%%%%%%%%%
\section*{ACKNOWLEDGEMENTS}
S.~T. thanks financial support for JSPS (No.~30318802).
%%%%%%%%%%%%%%%%%%%%%%%%%%%%%%%%


\begin{thebibliography}{10}

\bibitem{Perl}
S.~Perlmutter {\it et al.},
%``Measurements of Omega and Lambda
%from 42 High-Redshift Supernovae,''
Astrophys.\ J.\  {\bf 517}, 565 (1999);
A.~G.~Riess {\it et al.},
%``Observational Evidence from Supernovae for
%an Accelerating Universe and a Cosmological Constant,''
Astron.\ J.\  {\bf 116}, 1009 (1998);
Astron.\ J.\  {\bf 117}, 707 (1999).

\bibitem{review}
V.~Sahni and A.~A.~Starobinsky, Int.\ J.\ Mod.\ Phys.\ D \textbf{9},
373 (2000); V.~Sahni, Lect.\ Notes Phys.\ {} \textbf{653}, 141 (2004);
S.~M.~Carroll, Living Rev.\ Rel.\ {} \textbf{4}, 1 (2001);
T.~Padmanabhan, Phys.\ Rept.\ {} \textbf{380}, 235 (2003);
P.~J.~E.~Peebles and B.~Ratra, Rev.\ Mod.\
Phys.\ {} \textbf{75}, 559 (2003); S.~Nojiri and S.~D.~Odintsov,
Int.\ J.\ Geom.\ Meth.\ Mod.\ Phys.\  {\bf 4}, 115 (2007).

\bibitem{CST}
E.~J.~Copeland, M.~Sami and S.~Tsujikawa,
%``Dynamics of dark energy,''
Int.\ J.\ Mod.\ Phys.\  D {\bf 15}, 1753 (2006).

\bibitem{Capo}
S.~Capozziello, Int. J. Mod. Phys. {\bf D 11}, 483, (2002);
S.~Capozziello, V.~F.~Cardone, S.~Carloni and A.~Troisi,
Int. J. Mod. Phys. {\bf D}, 12, 1969 (2003);
S.~M.~Carroll, V.~Duvvuri, M.~Trodden and M.~S.~Turner,
%``Is cosmic speed-up due to new gravitational physics?,''
Phys.\ Rev.\  D {\bf 70}, 043528 (2004);
S.~Nojiri and S.~D.~Odintsov, Phys.\ Rev.\ D
\textbf{68}, 123512 (2003).

\bibitem{stensor}
L.~Amendola,
%``Scaling solutions in general non-minimal coupling theories,''
Phys.\ Rev.\  D {\bf 60}, 043501 (1999);
J.~P.~Uzan,
%``Cosmological scaling solutions of non-minimally
%coupled scalar fields,''
Phys.\ Rev.\  D {\bf 59}, 123510 (1999);
T.~Chiba,
%``Quintessence, the gravitational constant, and gravity,''
Phys.\ Rev.\ D {\bf 60}, 083508 (1999);
N.~Bartolo and M.~Pietroni,
%``Scalar tensor gravity and quintessence,''
Phys.\ Rev.\ D {\bf 61} 023518 (2000);
F.~Perrotta, C.~Baccigalupi and S.~Matarrese,
%``Extended quintessence,''
Phys.\ Rev.\ D {\bf 61}, 023507 (2000).

\bibitem{brane}
G.~R.~Dvali, G.~Gabadadze and M.~Porrati,
%``4D gravity on a brane in 5D Minkowski space,''
Phys.\ Lett.\ B {\bf 485}, 208 (2000).

\bibitem{quin}
Y.~Fujii, Phys.\ Rev.\ D {\bf 26}, 2580 (1982);
L.~H.~Ford,
%``Cosmological Constant Damping
%By Unstable Scalar Fields,''
Phys.\ Rev.\ D {\bf 35}, 2339 (1987);
C.~Wetterich, Nucl. \ Phys \ B. {\bf 302},
668 (1988);
B.~Ratra and J.~Peebles,
Phys. \ Rev \ D {\bf 37}, 321 (1988);
R.~R.~Caldwell, R.~Dave and P.~J.~Steinhardt,
%``Cosmological Imprint of an Energy Component
%with General Equation-of-State,''
Phys.\ Rev.\ Lett.\  {\bf 80}, 1582 (1998).

\bibitem{kes}
T.~Chiba, T.~Okabe and M.~Yamaguchi,
%``Kinetically driven quintessence,''
Phys.\ Rev.\  D {\bf 62}, 023511 (2000);
C.~Armendariz-Picon, V.~F.~Mukhanov and P.~J.~Steinhardt,
%``A dynamical solution to the problem of a small
%cosmological constant  and late-time cosmic acceleration,''
Phys.\ Rev.\ Lett.\  {\bf 85}, 4438 (2000).

\bibitem{sta}
A.~D.~Dolgov and M.~Kawasaki,
%``Can modified gravity explain accelerated cosmic expansion?,''
Phys.\ Lett.\  B {\bf 573}, 1 (2003);
V.~Faraoni,
%``Modified gravity and the stability of de Sitter space,''
Phys.\ Rev.\  D {\bf 72}, 061501 (2005);
S.~M.~Carroll, I.~Sawicki, A.~Silvestri and M.~Trodden,
%``Modified-Source Gravity and Cosmological Structure Formation,''
New J.\ Phys.\  \textbf{8}, 323 (2006);
R.~Bean, D.~Bernat, L.~Pogosian, A.~Silvestri and M.~Trodden,
%``Dynamics of Linear Perturbations in f(R) Gravity,''
Phys.\ Rev.\  D \textbf{75}, 064020 (2007);
Y.~S.~Song, W.~Hu and I.~Sawicki,
%``The large scale structure of f(R) gravity,''
Phys.\ Rev.\  D {\bf 75}, 044004 (2007);
T.~Faulkner, M.~Tegmark, E.~F.~Bunn and Y.~Mao,
%``Constraining f(R) gravity as a scalar tensor theory,''
Phys.\ Rev.\  D {\bf 76}, 063505 (2007).

\bibitem{APT}
L.~Amendola, D.~Polarski and S.~Tsujikawa,
%``Are f(R) dark energy models cosmologically viable ?,''
Phys.\ Rev.\ Lett.\  {\bf 98}, 131302 (2007);
Int.\ J.\ Mod.\ Phys.\  D {\bf 16}, 1555 (2007).

\bibitem{lgc}
G.~J.~Olmo,
Phys.\ Rev.\  D {\bf 72}, 083505 (2005);
A.~L.~Erickcek, T.~L.~Smith and M.~Kamionkowski,
Phys.\ Rev.\  D {\bf 74}, 121501 (2006);
V.~Faraoni,
Phys.\ Rev.\  D {\bf 74}, 023529 (2006);
T.~Chiba, T.~L.~Smith and A.~L.~Erickcek,
%``Solar System constraints to general f(R) gravity,''
Phys.\ Rev.\  D {\bf 75}, 124014 (2007);
I.~Navarro and K.~Van Acoleyen,
%``f(R) actions, cosmic acceleration and local tests of gravity,''
JCAP {\bf 0702}, 022 (2007).

\bibitem{AGPT}
L.~Amendola, R.~Gannouji, D.~Polarski and S.~Tsujikawa,
%``Conditions for the cosmological viability of f(R) dark energy models,''
Phys.\ Rev.\  D {\bf 75}, 083504 (2007).

\bibitem{Li}
B.~Li and J.~D.~Barrow,
%``The Cosmology of f(R) Gravity in Metric Variational Approach,''
Phys.\ Rev.\  D {\bf 75}, 084010 (2007).

\bibitem{AT08}
L.~Amendola and S.~Tsujikawa,
%``Phantom crossing, equation-of-state singularities, and local gravity
%constraints in $f(R)$ models,''
Phys.\ Lett.\  B {\bf 660}, 125 (2008).

\bibitem{Hu}
W.~Hu and I.~Sawicki,
%``Models of f(R) Cosmic Acceleration that Evade Solar-System Tests,''
Phys.\ Rev.\  D {\bf 76}, 064004 (2007).

\bibitem{Star}
A.~A.~Starobinsky,
%``Disappearing cosmological constant in f(R) gravity,''
JETP Lett.\  {\bf 86}, 157 (2007).

\bibitem{Appleby}
S.~A.~Appleby and R.~A.~Battye,
%``Do consistent $F(R)$ models mimic General Relativity plus $\Lambda$?,''
Phys.\ Lett.\  B {\bf 654}, 7 (2007);
arXiv:0803.1081 [astro-ph].

\bibitem{Tsuji}
S.~Tsujikawa,
%``Observational signatures of f(R) dark energy models 
%that satisfy cosmological and local gravity constraints,''
Phys.\ Rev.\  D {\bf 77}, 023507 (2008).

\bibitem{NO07}
S.~Nojiri and S.~D.~Odintsov,
Phys.\ Lett.\  B {\bf 657}, 238 (2007);
G.~Cognola {\it et al.},
Phys.\ Rev.\  D {\bf 77}, 046009 (2008).

\bibitem{TUT}
S.~Tsujikawa, K.~Uddin and R.~Tavakol,
%``Density perturbations in f(R) gravity theories in metric and
%Palatini formalisms,''
Phys.\ Rev.\  D {\bf 77}, 043007 (2008).

\bibitem{CT}
S.~Capozziello and S.~Tsujikawa,
%``Solar system and equivalence principle constraints 
%on $f(R)$ gravity by chameleon approach,''
Phys.\ Rev.\  D {\bf 77}, 107501 (2008).

\bibitem{TUMTY}
S.~Tsujikawa, K.~Uddin, S.~Mizuno, R.~Tavakol and J.~Yokoyama,
%``Constraints on scalar-tensor models of dark energy from
%observational and local gravity tests,''
Phys.\ Rev.\  D {\bf 77}, 103009 (2008).

\bibitem{Bar}
M.~Bartelmann and P.~Schneider,
%``Weak Gravitational Lensing,''
Phys.\ Rept.\  {\bf 340}, 291 (2001);
D.~Munshi, P.~Valageas, L.~Van Waerbeke and A.~Heavens,
%``Cosmology with Weak Lensing Surveys,''
arXiv:astro-ph/0612667.

\bibitem{Bena}
K.~Benabed and F.~Bernardeau,
%``Testing quintessence models with large-scale structure growth,''
Phys.\ Rev.\  D {\bf 64}, 083501 (2001).

\bibitem{Jain}
B.~Jain and A.~Taylor,
%``Cross-correlation Tomography: Measuring Dark Energy 
%Evolution with Weak Lensing,''
Phys.\ Rev.\ Lett.\  {\bf 91}, 141302 (2003);
M.~Takada and B.~Jain,
%``Cosmological parameters from lensing power spectrum
%and bispectrum tomography,''
Mon.\ Not.\ Roy.\ Astron.\ Soc.\  {\bf 348}, 897 (2004);
Y.~S.~Song and L.~Knox,
%``Dark energy tomography,''
Phys.\ Rev.\  D {\bf 70}, 063510 (2004).

\bibitem{Simpson}
F.~Simpson and S.~Bridle,
%``Illuminating Dark Energy with Cosmic Shear,''
Phys.\ Rev.\  D {\bf 71}, 083501 (2005).

\bibitem{Heavens}
A.~F.~Heavens, T.~D.~Kitching and A.~N.~Taylor,
%``Measuring dark energy properties with 3D cosmic shear,''
Mon.\ Not.\ Roy.\ Astron.\ Soc.\  {\bf 373}, 105 (2006);
A.~N.~Taylor, T.~D.~Kitching, D.~J.~Bacon and A.~F.~Heavens,
%``Probing dark energy with the shear-ratio geometric test,''
Mon.\ Not.\ Roy.\ Astron.\ Soc.\  {\bf 374}, 1377 (2007).

\bibitem{Takada}
M.~Takada and S.~Bridle,
%``Probing dark energy with cluster counts and cosmic 
%shear power spectra: including the full covariance,''
New J.\ Phys.\  {\bf 9}, 446 (2007).

\bibitem{AKS}
L.~Amendola, M.~Kunz and D.~Sapone,
%``Measuring the dark side (with weak lensing),''
JCAP {\bf 0804}, 013 (2008).

\bibitem{Jaza}
B.~Jain and P.~Zhang,
%``Observational Tests of Modified Gravity,''
arXiv:0709.2375 [astro-ph].

\bibitem{Vivi07}
V.~Acquaviva and L.~Verde,
JCAP {\bf 0712}, 001 (2007).

\bibitem{Bean}
I.~Laszlo and R.~Bean,
%``Nonlinear growth in modified gravity theories of dark energy,''
Phys.\ Rev.\  D {\bf 77}, 024048 (2008).

\bibitem{Viviana}
V.~Acquaviva, C.~Baccigalupi and F.~Perrotta,
%``Weak lensing in generalized gravity theories,''
Phys.\ Rev.\  D {\bf 70}, 023515 (2004).

\bibitem{Uzan}
C.~Schimd, J.~P.~Uzan and A.~Riazuelo,
%``Weak lensing in scalar-tensor theories of gravity,''
Phys.\ Rev.\  D {\bf 71}, 083512 (2005).

\bibitem{Boi}
B.~Boisseau, G.~Esposito-Farese, D.~Polarski 
and A.~A.~Starobinsky,
%``Reconstruction of a scalar-tensor theory of gravity 
%in an accelerating universe,''
Phys.\ Rev.\ Lett.\  {\bf 85}, 2236 (2000).

\bibitem{Hwang}
J.~c.~Hwang and H.~Noh,
Phys.\ Rev.\  D {\bf 65}, 023512 (2002).

\bibitem{Tsuji07}
S.~Tsujikawa,
%``Matter density perturbations and effective 
%gravitational constant in modified gravity models of dark energy,''
Phys.\ Rev.\  D {\bf 76}, 023514 (2007).

\bibitem{KM}
K.~Koyama and R.~Maartens,
%``Structure formation in the DGP cosmological model,''
JCAP {\bf 0601}, 016 (2006).

\bibitem{Brans}
C.~Brans and R.~H.~Dicke,
%``Mach's principle and a relativistic theory of gravitation,''
Phys.\ Rev.\  {\bf 124}, 925 (1961).

\bibitem{Chiba03}
T.~Chiba,
%``1/R gravity and scalar-tensor gravity,''
Phys.\ Lett.\  B {\bf 575}, 1 (2003).

\bibitem{lgccon}
C.~D.~Hoyle {\it et al.,}
%``Sub-millimeter tests of the gravitational inverse-square law,''
Phys.\ Rev.\ D \textbf{70}, 042004 (2004).

\bibitem{KW}
J.~Khoury and A.~Weltman,
Phys.\ Rev.\ Lett.\  {\bf 93}, 171104 (2004).

\bibitem{Dodelson}
S.~Dodelson, {\it Modern Cosmology}, Academic Press (2003).

\bibitem{BBKS}
J.~M.~Bardeen, J.~R.~Bond, N.~Kaiser and A.~S.~Szalay,
%``The Statistics Of Peaks Of Gaussian Random Fields,''
Astrophys.\ J.\  {\bf 304}, 15 (1986).

\bibitem{halo}
R.~E.~Smith {\it et al.}, Mon.\,Not.\,Roy.\,Astron.\,Soc.\,
{\bf 341}, 1311 (2003).

\end{thebibliography}
\end{document}